\documentclass[twocolumn,showpacs,preprintnumbers,amsmath,amssymb,superscriptaddress]{revtex4-1}

\usepackage{graphicx}
\usepackage{dcolumn}
\usepackage{bm}
\usepackage{natbib}
\usepackage[usenames,dvipsnames]{xcolor}
\usepackage{pifont}
\usepackage{relsize}
\definecolor{lightgreen}{rgb}{0,1,0}
\definecolor{darkgray}{gray}{0.20}
\usepackage{subfigure}
\usepackage{hyperref}
\usepackage{float} 

\begin{document}

\title{The Rise of China in the International Trade Network: \\
A Community Core Detection Approach}

\author{Zhen Zhu}\altaffiliation{Corresponding author. \\ Email: \href{mailto:zhen.zhu@imtlucca.it}{zhen.zhu@imtlucca.it}.}
	  \affiliation{IMT Institute 
  for Advanced Studies Lucca, Piazza S. Ponziano 6, 55100 Lucca, Italy}   
\author{Federica Cerina} \affiliation{Department of Physics,
  Universit\`{a} degli Studi di Cagliari, Cagliari, Italy}\affiliation{Linkalab, Complex
  Systems Computational Laboratory, Cagliari 09129, Italy} 
\author{Alessandro Chessa}  \affiliation{IMT Institute 
  for Advanced Studies Lucca, Piazza S. Ponziano 6, 55100 Lucca, Italy}\affiliation{Linkalab, Complex
  Systems Computational Laboratory, Cagliari 09129, Italy}      
\author{Guido Caldarelli}\affiliation{IMT Institute 
  for Advanced Studies Lucca, Piazza S. Ponziano 6, 55100 Lucca, Italy} \affiliation{Linkalab, Complex
  Systems Computational Laboratory, Cagliari 09129, Italy} \affiliation{ISC-CNR, Via dei Taurini 19 00185 Rome, Italy} \affiliation{London Institute for Mathematical Sciences, 35a South Street Mayfair, London,
United Kingdom W1K 2XF} 
\author{Massimo Riccaboni} \affiliation{IMT Institute 
  for Advanced Studies Lucca, Piazza S. Ponziano 6, 55100 Lucca, Italy} \affiliation{DMSI, KU Leuven, Belgium}

\begin{abstract}
Theory of complex networks proved successful in the description of a variety of static networks ranging from biology to computer and social sciences and to economics and finance. 
Here we use network models to describe the evolution of a particular economic system, namely the International Trade Network 
(ITN). Previous studies often assume that globalization and regionalization in international trade are contradictory to each other. We re-examine the relationship between globalization and regionalization 
by viewing the international trade system as an interdependent complex network. We use the modularity optimization method to detect communities and community cores in the ITN during the 
years 1995-2011. We find rich dynamics over time both inter- and intra-communities. Most importantly, we have a multilevel description of the evolution where the global dynamics (i.e., communities disappear or reemerge) 
tend to be correlated with the regional dynamics (i.e., community core changes between community members). In particular, the Asia-Oceania community disappeared and reemerged over 
time along with a switch in leadership from Japan to China. Moreover, simulation results show that the global dynamics can be generated by a preferential attachment mechanism both inter- and intra-
communities.
\end{abstract}

\maketitle 

\noindent {\it ``Befriend a distant state while attacking a neighbor.''} 
\begin{flushright}
{\it Thirty-Six Stratagems}
\end{flushright}

\section{Introduction} \label{sec:intro}

Complex networks are a modern way to characterize mathematically a series of different systems in the shape of subunits (nodes) connected by their interaction (edges)\cite{ABRMP}. Such modeling has been proved to be fruitful for the description 
of a variety of different phenomena ranging from biology\cite{GCbook2} to social sciences\cite{SIAM}, economics\cite{kitsak2010,chessa2013} and 
finance\cite{nature}. Here we move forward by considering the change in shape of some topological quantities (namely the community structure) during the evolution of a particular instance of complex network.
Such instance is represented by the International Trade Network (ITN), a structure composed by the various world nations, connected by international trade. 

The last two decades have witnessed both intensified globalization and regionalization in international trade. The former is evidenced by the formation of unbiased trade relationships across diverse groups of countries while the latter 
is evidenced by the formation of regional trade agreements and free trade areas. When empirically testing the above two phenomena, previous studies often assume that they are contradictory to each other and try to answer questions like 
``Has the world become more globalized or regionalized?'' Based on various data sets and methodologies, some studies conclude with strong evidence of globalization~\cite{hummels2007transportation}, while others argue the 
opposite~\cite{edward1999new,chortareas2004trade}, while yet others have mixed results~\cite{arribas2009measuring}.

A fast-growing literature has been built in recent years by viewing the international trade system as an interdependent complex network, 
where countries are represented by nodes and trade relationships are represented by edges~\cite{serrano2003topology,garlaschelli2005structure,fagiolo2009world,riccaboni2010,de2011world,riccaboni2013}. 
As a result, many topics in international economics have been re-investigated through the lens of networks, and globalization and regionalization are certainly no exception. 
However, even with the networks approach, the question of whether we have a more globalized or regionalized world is still answered with mixed results~\cite{kim2002longitudinal,tzekina2008evolution,piccardi2012existence,reyes2014regional}.
Moreover, the contribution of network analysis to our understanding of international trade has been questioned, since there is still little evidence about the importance of global effects on the performances of single countries (nodes) 
and trade linkages.

In this paper, we re-examine the relationship between globalization and regionalization from a different angle. 
Instead of assuming that the two are contradictory to each other and attempting to figure out which is dominating the other, we take into account the dynamics 
in the ITN at both regional level and global level and investigate the interaction between the two. Besides that, we will take advantage of a unique ``natural experiment'', that is the opening of China to the world
trade and the entry of China in the World Trade Organization in 2001, to analyze the reverberations of a huge country-specific shock on the structure of the ITN.

We make use of the CEPII BACI Database~\cite{gaulier2010baci} to build up the ITN and use the modularity optimization method~\cite{newman2004finding} to detect both communities 
and community cores in the ITN during the years 1995-2011. The global dynamics can be seen if communities disappear or reemerge over time and the regional dynamics can be seen if leadership (community core) changes between community members.   

We find that the global dynamics tend to be correlated with the regional dynamics. In particular, the Asia-Oceania community displayed an interesting interaction between the two, which can be roughly summarized in the following three stages:
\begin{enumerate}
\item During 1995-2001, the Asia-Oceania community was present\footnote{Only with a brief interruption in 1998, when the Asia-Oceania community was integrated with the America community. Also, during 1999-2001, while China was always 
a member of the Asia-Oceania community, Japan, Oceania, part of the Southeast Asia, and some other Asian economies were integrated with the America community.} in the ITN and was led 
by Japan\footnote{During 1999-2001, when Japan was integrated with America, the Asia-Oceania community was led by Hong Kong instead.}; 
\item During 2002-2004, the Asia-Oceania community disappeared and was integrated with the American community, which was led by the United States;
\item During 2005-2011, the Asia-Oceania community reemerged and was led by China. 
\end{enumerate} 

Our simulation results show that the disappearance and reemergence of the communities can be generated by a preferential attachment mechanism both inter- and intra-communities. 
Furthermore, the rise of China in the Asia-Oceania community can be explained by its dramatic increase of inter-community trade since 2002. 
The intuition is that, the Asia-Oceania community collapsed after China entered the WTO and built strong trade relationships with other communities, 
especially with the external cores, i.e., the United States and Germany, and China became regionally attractive and restored the Asia-Oceania community and emerged as the community leader after it 
gained a significant portion of trade globally. These can be considered as a series of strategic moves implemented by China's foreign trade policy. 
As quoted in the beginning of the paper, a classical stratagem to achieve regional power is to befriend a distant state. 

Our contribution to the analysis of the ITN is twofold.
First, we provide some evidence of a clear violation of the Barab\'{a}si-Albert preferential attachment rule~\citep{ABRMP} and the law of gravity in the world trade. Second, we identify a mechanism that can account for this departure from the gravity law
and validate it via simulations, historical reconstruction and empirical analysis. We show that by increasing its global export China is also increasing the chance to import more goods from regional trading partners. In other words,
part of the Chinese export growth shock gets transmitted to other economies in the same region by means of a corresponding increase in Chinese imports of intermediate goods and partial delocalization of production. The transmission 
mechanism we identify provides further support for a network approach to the analysis of world trade, since we show how local changes in the intensity of trade diffuse to other nodes in the network. We argue that a reductionist approach, 
which relies exclusively on node and link specific information misses some important network effects in the world trade structure. Even though we limit our analysis to the Chinese case, a similar argument applies to emergence of the Arabic
community after 9/11 and other relevant shocks to the world trade structure.

The rest of the paper is structured as follows. Section \ref{sec:method} describes our methodology of community detection and community core detection, 
respectively. Section \ref{sec:data} summarizes the data we use to build the ITN. The detection results are reported and discussed in Section \ref{sec:results}. 
A model and its simulation results and some empirical evidence to explain the dynamics observed are presented in Section \ref{sec:exp}. 
Finally, Section \ref{sec:conclusion} concludes the paper.

\section{Methodology}\label{sec:method}

\subsection{Community Detection}

It is well known that one of the main features of networks is community structure, i.e. their capacity to organize nodes in clusters, with many edges connecting nodes in the same cluster 
and few connecting nodes between different ones. Detecting communities is of great importance in various disciplines where systems can be mapped onto networks.

In the following we use the modularity optimization method introduced by Newman and Girvan~\cite{newman2004finding}. It is based on the idea that a random graph is not 
expected to have a cluster structure, so the possible existence of clusters is revealed by the comparison between the actual
density of edges in a subgraph and the density one would
expect to have in the subgraph if the nodes of the graph
were attached regardless of community structure. This
expected edge density depends on the chosen
null model, i.e., a copy of the original graph keeping some of its
structural properties but without community structure~\cite{fortunato2010}.

The most popular null model, introduced by Newman and Girvan, keeps the degree sequence and consists of a randomized version of the 
original graph, where edges are rewired at random, under the constraint that the expected degree of each node matches the degree of the node in the original graph~\cite{newman2004finding}. 

The modularity function to be optimized is, then, defined as~\cite{newman2004finding}:
\begin{equation}
Q = \frac{1}{2m} \sum_{ij} (A_{ij}-P_{ij})\delta({C_i,C_j}) 
\end{equation}

\noindent where the summation operator runs over all the node pairs, $A$ is the adjacency matrix, $m$ is the total number of edges and is the 
expected number of edges between the nodes $i$ e $j$ for a given null model. The $\delta$ function will result in a null contribution for couples of nodes not belonging to the same community ($C_i\neq C_j$). For an unweighted
network, the choice $P_{ij} = \frac{k_i k_j}{2m}$ is to take a random network with the same degree sequence as the original one.

This method suffers from various problems, the most important one being the existence of a resolution limit~\cite{fortunato2012}, which prevents it 
from detecting smaller modules. However, it is by far the most used community detection method. It delivers good results and has some nice features such as being a global criterion and simple to implement.

\subsection{Community Core Detection}

The main problem of all algorithms for community detection is the fact that the community definition does not provide any information about the importance of any individual node inside the community. Nodes of a community
do not have the same importance for the community stability: the removal of a node in the ``core'' of a network
affects the partition much more than the deletion of a node that stays on the periphery of the community~\cite{deleo2013}. Therefore, in the following we 
complement community detection with a novel way of detecting cores inside communities by using the properties of the modularity
function.

By definition, if the modularity associated with a network has been optimized, every perturbation in the
partition leads to a negative variation in the modularity ($\mathrm{d}Q$). If we move a node from a partition,
we have $M-1$ possible choices (with $M$ as the number of communities) as the node's new host community. It is possible to define 
the $|\mathrm{d}Q|$ associated with each node as the smallest variation in absolute value (or the closest to 0 since $\mathrm{d}Q$ is always 
a negative number) for all the possible choices. This is a measure of how important that node is to its community~\cite{deleo2013}.
 
It follows that, within a community, the node with the highest $|\mathrm{d}Q|$ is the most important one and it can be reasoned as the leader of that 
community, in terms of the strength of intra-community edges. To also take into account the overall centrality of the node, a better indicator of leadership 
would be $|\mathrm{d}Q|*strength$, where $strength$ is simply the node strength in the network~\cite{cerina2014}. 

Finally, in order to have a better visualization of the relative importance of nodes in different communities we use the $CS$ index, ranging from 0 to 1, which is simply $|\mathrm{d}Q|*strength$ normalized for each community. 

\section{Data} \label{sec:data}

We use the BACI database~\cite{gaulier2010baci} to build up the ITN. BACI is the world trade database developed by the CEPII at a high level of product disaggregation. 
Original data are provided by the United Nations Statistical Division (COMTRADE database). BACI is constructed using an original procedure that reconciles the declarations of the 
exporter and the importer. This harmonization procedure considerably extends the number of countries for which trade data are available, as compared to the original COMTRADE. 
Furthermore, BACI provides bilateral values and quantities of exports at the HS 6-digit product level, for more than 200 countries since 
1995.\footnote{See the CEPII website, \url{http://www.cepii.fr/CEPII/en/bdd_modele/presentation.asp?id=1}, for further information about BACI.}  

We use the BACI database from 1995 to 2011 and, for each year, we sum up all the bilateral commodity flows between any two countries. We construct the 
ITN with countries as nodes and with the total bilateral trade flow between countries $i$ and $j$ as the edge weight $w_{ij}$.

\section{Detection Results} \label{sec:results}

\subsection{Global Dynamics versus Regional Dynamics}

During the years 1995-2011 we have examined, the ITN was mainly characterized by three communities, namely, the America community, the Europe community, 
and the Asia-Oceania community. According to the United Nations definitions of macro geographical regions\footnote{See the website of the United Nations Statistics 
Division, \url{https://unstats.un.org/unsd/methods/m49/m49regin.htm}.}, the America community is more or less comprised of Americas. The Europe community is more or
less comprised of Europe and Central Asia. The Asia-Oceania community is more or less comprised of Eastern Asia, Southern Asia, South-Eastern Asia, and 
Oceania.\footnote{Countries in Africa and Western Asia don't have consistent community memberships over time. Therefore, they are not classified in any of the three communities.} 

However, among the three main communities, the America community and the Europe community were more stable than the Asia-Oceania community. First, over the 17 years, 
the America community and the Europe community were always present while the Asia-Oceania community experienced disappearance and reemergence. Second, the intra-community 
structure was more stable in the America community and the Europe community in a sense that the community leaders (cores) over time were always the United States and Germany, 
respectively. The Asia-Oceania community on the other hand experienced a leadership change from Japan to China. 
   
Because the Asia-Oceania community has shown rich dynamics both internally and externally, in Subsection \ref{subsec:asia} we focus our attention on it.

\subsection{The Asia-Oceania Community}\label{subsec:asia}

As mentioned in Section \ref{sec:intro}, the dynamics of the Asia-Oceania community can be roughly divided into three stages, namely, its presence 
with Japan's leadership during 1995-2001, its disappearance and integration with the America community during 2002-2004, and finally its reemergence with China's leadership during 2005-2011. 

\begin{figure}[!t]
\centering
{\includegraphics[width=1\columnwidth]{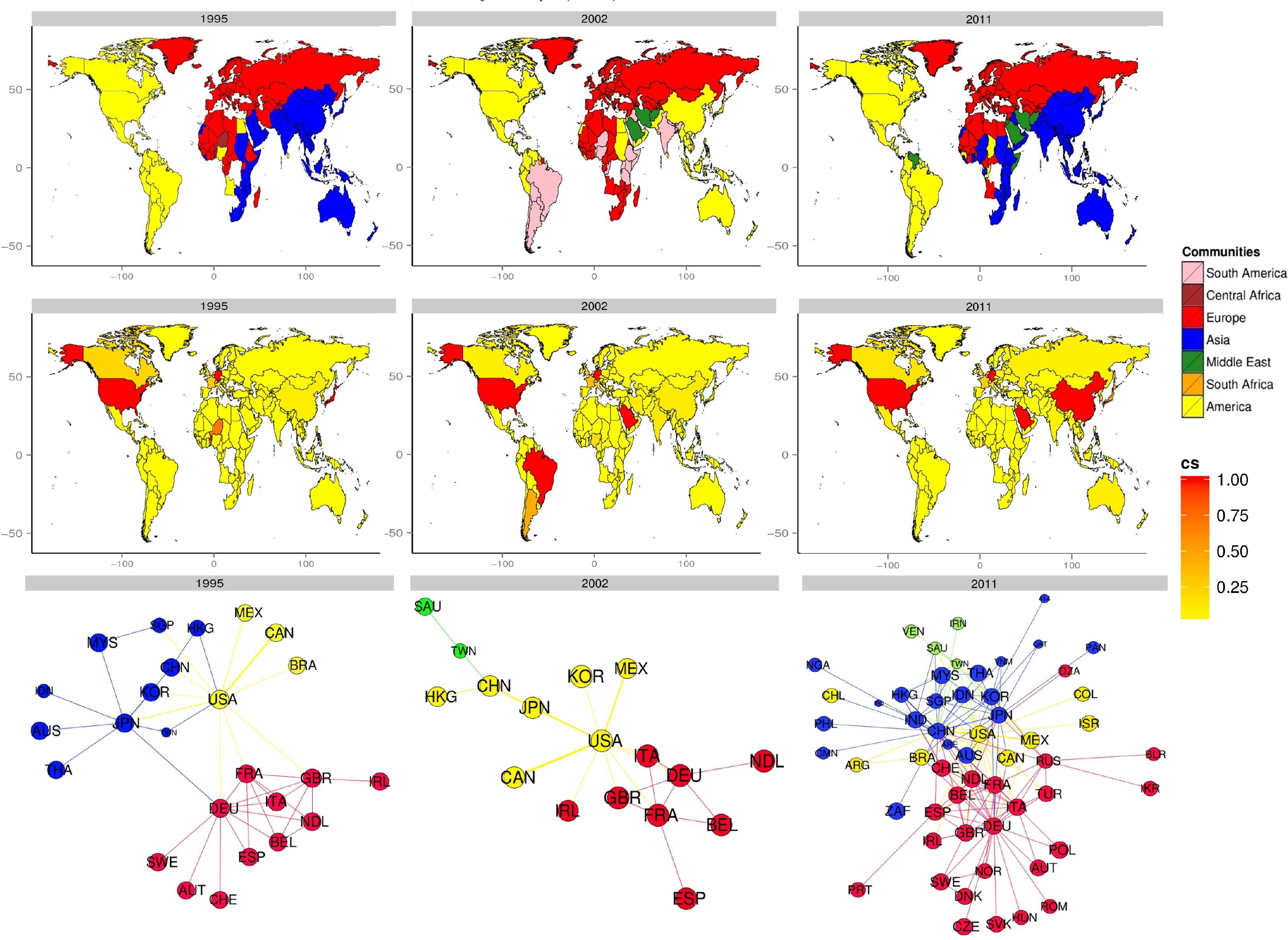}}
\caption{{\bf Community and Community Core Detection Results.} From left to right, the three columns are corresponding to the years 1995, 2002, and 2011, respectively. The first row shows the Newman-Girvan community detection results. The America community is colored yellow, the Europe community is colored red, and the Asia-Oceania 
community is colored blue. Asia-Oceania and America were separated from each other in 1995 and 2011 but was integrated in 2002. The second row shows the community core 
detection results by normalizing $|\mathrm{d}Q|*strength$ for each community. The redness of each country is proportional to its relative magnitude of $|\mathrm{d}Q|*strength$ within its community. The reddest country in the Asia-Oceania community 
was Japan back in 1995 but became China in 2011. Finally, the third row provides a topological view of the community structure in the three years. Again, Japan was central in the Asia-Oceania community in 1995 and it was replaced 
by China in 2011.} \label{fig1}
\end{figure}

The same pattern is shown in Figure \ref{fig1}, where three years, 1995, 2002, and 2011, are selected to represent the three stages respectively.\footnote{The results for all years from 1995 to 2011 are in the Appendix.} 
The first row shows the community maps in the three years. The America community is colored yellow, the Europe community is colored red, and the Asia-Oceania 
community is colored blue. Notice that in 2002 the blue community was by and large merged with the yellow community.\footnote{As discussed in Section \ref{sec:intro}, 
another interesting change in the world trade community structure is the emergence of the Arab community after 2001. This interesting phenomenon deserves further scrutiny in future research.} The second row shows the community core 
detection results for the three years. The redder the more important the country is in reserving its community. Equivalently, the yellower the less important the 
country is in reserving its community. This can be used to identify the leaders in the communities. Notice that in 1995 the reddest country in the Asia-Oceania community 
was Japan while in 2011 China became the reddest. Finally, the third row provides a topological view of the community structure in the three years. Again, Japan was central in the Asia-Oceania community in 1995 and it was replaced 
by China in 2011.   

\section{Explanations for the Dynamics in the Asia-Oceania Community} \label{sec:exp}

Given its breathtaking economic growth during 1995-2011, it is not surprising to see China's rise in the regional trade 
community. The rationale behind is the long-established gravity model of trade~\cite{bergstrand1985gravity,baldwin2006gravity,carrere2006revisiting}. That is, 
the increased economic mass of China tends to attract more trade flows with other economies. What remains unexplained, however, is the fact that the leadership change 
from Japan to China is correlated with the disappearance and reemergence of the Asia-Oceania community.

To address the linkage between the global dynamics and the regional dynamics, we propose a model with weight-driven preferential attachment both inter- and intra-communities.    

\subsection{A Model with Inter- and Intra-Communities Preferential Attachment}

Since the number of countries in the ITN is constant over time and the evolution of the ITN is only concerned with the trade flows between countries, our model is 
therefore based on a fixed number of nodes and a weight-driven preferential attachment mechanism both inter- and intra-communities.\footnote{There exists some related 
literature to our model. For example, Barrat et al.~\cite{barrat2004weighted} and Riccaboni and Schiavo~\cite{riccaboni2010} examine the network evolution with dynamic edge weights. Li and Maini~\cite{li2005evolving} 
investigate the network properties with preferential attachment both inter- and intra-communities. However, to the best of our knowledge, our model is the first attempt to 
bring the preferential attachment mechanism both inter- and intra-communities to the context of a weighted network with a fixed number of nodes.} Additionally, our model is 
based on an undirected network because the ITN is constructed by total bilateral trade flows. 

The initial status of the network is characterized by $M$ arbitrarily imposed communities.\footnote{In the context of ITN, the communities can be formed, for instance, by continents.} 
For simplicity, each community has the same number of nodes, $m_0$. As a subgraph, each community is completely connected with a equal edge weight, i.e., every node is connected with 
every node by the same edge weight in the community. Between any two communities, there is only one edge connecting two randomly selected nodes in the two communities respectively. 
Again for simplicity, the inter-community edge weight is set to equal the initial intra-community edge weight. After the initial set-up, each period the preferential attachment mechanism is comprised of the following steps:
\begin{enumerate}
\item One node, $i$, is selected based on a uniform distribution across all the nodes in the network;
\item Suppose that $i$ belongs to community $j$, by chance, $i$ can increase its edge weight with a node outside community $j$. And the reach-out probability is:
\begin{equation}
R^{inter} = \frac{s_{i,j}^{intra}}{\alpha\sum_{k} s_{k,j}^{intra}}
\end{equation}
where $s_{i,j}^{intra}$ is the intra-community strength of node $i$ in community $j$, i.e., the sum of the edge weights between node $i$ and all other members in 
community $j$. $\alpha\geq1$ and a big $\alpha$ means that any node will have low probability to reach out to other communities. The intuition is that, it is difficult 
for a node to reach out given a big $\alpha$ and\footnote{In the context of the ITN, a high value of $\alpha$ can be interpreted as trade barriers such as tariffs, 
transportation costs, and language difference.}, within a community, the nodes with more intra-community strength are more likely to reach out;
\item There are $(M-1)m_0$ nodes outside community $j$. The one with which $i$ increases the edge weight is determined by the following probability mass function:
\begin{equation}
P_{u,-j}^{inter} = \frac{s_{u,-j}^{intra}}{\sum_{-j}\sum_{u} s_{u,-j}^{intra}}
\end{equation} 
where $-j$ is a community other than community $j$. The intuition is that, if $i$ is able to reach out, it will prefer to reach out to the ones with more intra-community 
strength in their own communities. After the inter-community node is identified, the edge weight between it and $i$ will be increased by $\beta^{inter}$;
\item The next step for $i$ is to choose a neighbor in the same community $j$ to increase the edge weight. The one is selected by the following probability mass function:
\begin{equation}
P_{-i,j}^{intra} = \frac{(1-\gamma)s_{-i,j}^{intra}+\gamma\sum_{-j} s_{-i,-j}^{inter}}{(1-\gamma)\sum_{-i} s_{-i,j}^{intra} + \gamma\sum_{-j}\sum_{-i} s_{-i,-j}^{inter}}
\end{equation}
where $-i$ is a neighbor to $i$ in the community $j$. $0\leq\gamma\leq1$ and when $\gamma$ gets close to 1, although $i$ prefers to increase the edge weight with the neighbors 
with more intra-community strength, it prefers even more the ones with more inter-community strength. After the neighbor is identified, the edge weight between it and $i$ will be increased by $\beta^{intra}$;
\item Finally, the modularity optimization method is used to detect the community structure, which may deviate from the original set-up.   
\end{enumerate} 

\subsection{Simulation Results} 

The initial status of our simulation is a network with 3 preset communities. Each community has 5 nodes and, as mentioned above, each community is completely connected and 
there is a single edge between any two communities. Other model parameters are $\alpha=10$, $\beta^{intra}=0.05$, $\beta^{inter}=2$, and $\gamma=0.9$, respectively. 
Setting $alpha$ to 10 and having a relatively big $\beta^{inter}$ compared to $\beta^{intra}$ are to make it difficult for a node to reach out to other communities so that 
the preset community structure can be restored over time. However, when a node does reach out, it is enough to introduce a perturbation to the community structure. We have run the above mentioned preferential attachment mechanism for 5000 periods. 

\begin{figure}[!t]
\centering
{\includegraphics[width=1\columnwidth]{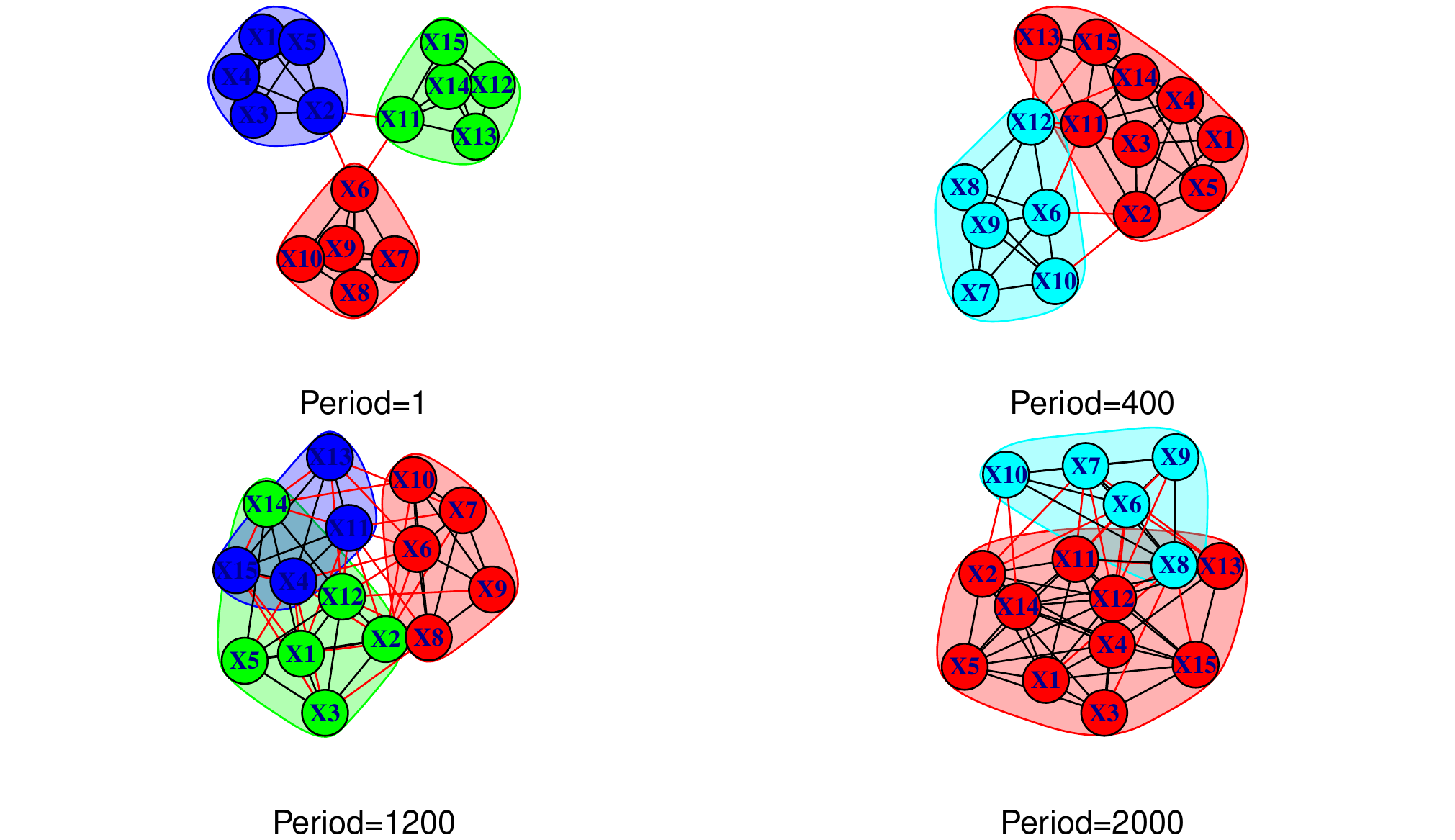}}
\caption{{\bf Simulation Results.} The simulation is based on a preferential attachment mechanism both inter- and intra-communities. The model parameters are $\alpha=10$, $\beta^{intra}=0.05$, $\beta^{inter}=2$, and $\gamma=0.9$, respectively. Different colors represent different communities detected by the Newman-Girvan method. The inter-community edges are colored red while the intra-community ones are colored black. Although the community detection takes into account the edge weights, all the edges in the figure have the same width. In period 1, three predetermined communities, X1-X5, X6-X10, and X11-X15, are imposed in the network. The number of communities detected in this 15-node network bounces back and forth between 3 and 2 during the simulated periods. That is, like what we observe from the ITN, the 
disappearance and reemergence of the communities can be generated by the preferential attachment mechanism both inter- and intra-communities.} \label{fig2}
\end{figure}

Figure \ref{fig2} selects 4 periods of our simulation. The 3 preset communities are X1-X5, X6-X10, and X11-X15, respectively. Different colors represent different communities 
detected by the modularity optimization method. The red edges are inter-community ones while the black ones are intra-community. Like what we observe from the ITN, the 
disappearance and reemergence of the communities can be generated by the preferential attachment mechanism both inter- and intra-communities. In fact, the number of communities detected in this 15-node network bounces back and forth between 3 and 2 during the simulated periods.    

\subsection{Empirical Evidence}

We now turn back to the ITN and present some empirical evidence for the preferential attachment mechanism both inter- and intra-communities. 

\begin{figure}[!t]
\centering
{\includegraphics[width=1\columnwidth]{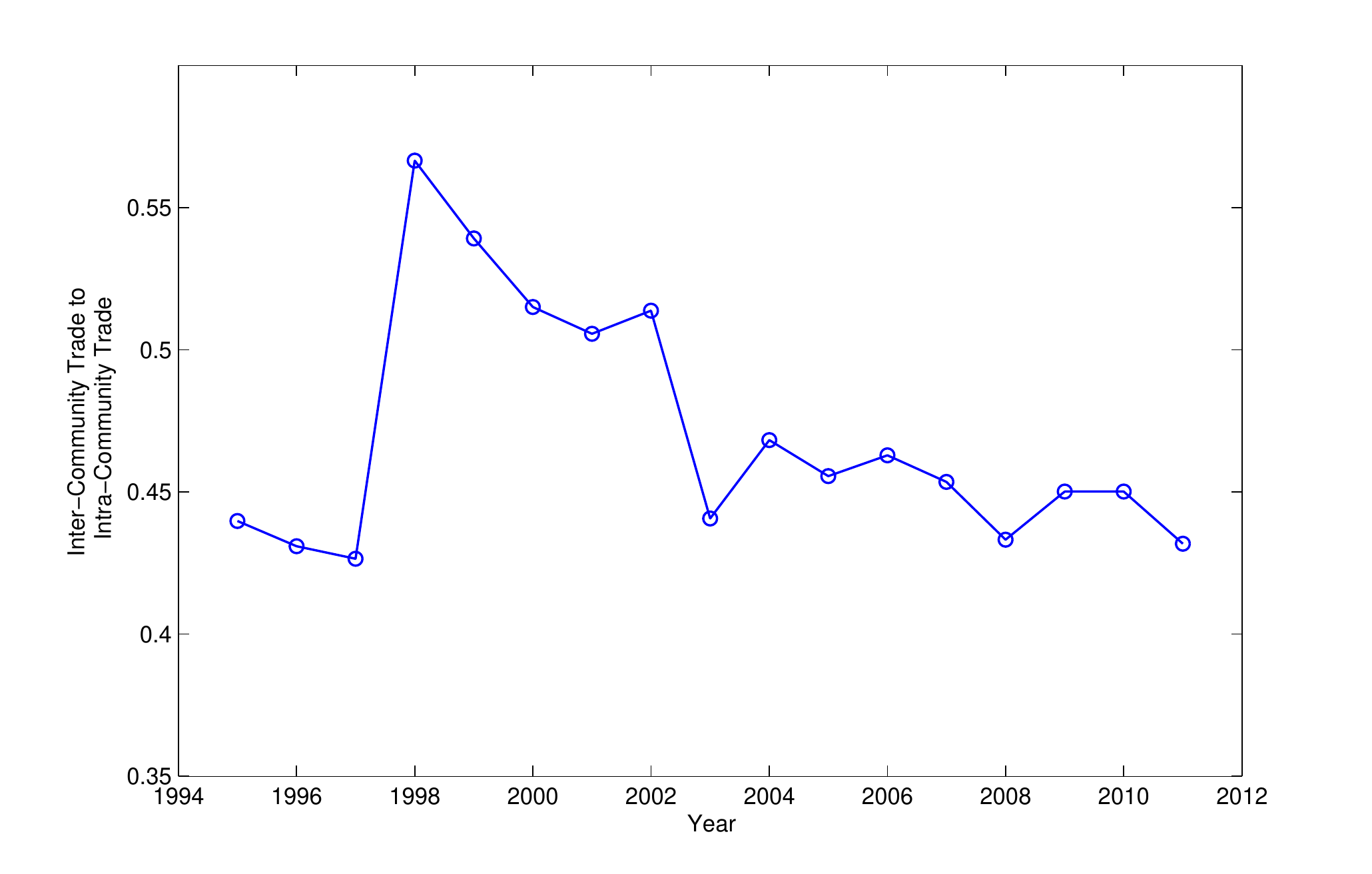}}
\caption{{\bf Inter- versus Intra-Community Trade Ratio between Asia-Oceania and America.} We calculate the ratio of the inter-community trade to the intra-community trade between the Asia-Oceania 
community and the America community. The ratio first went up and then went down and formed a hump shape over time. This finding coincides with the disappearance and reemergence of the Asia-Oceania community observed in Figure \ref{fig1}. } \label{fig3}
\end{figure}

First, for the inter-community dynamics, we calculate the ratio of the inter-community trade to the intra-community trade between the Asia-Oceania 
community and the America community. As shown in Figure \ref{fig3}, the ratio first went up and then went down and formed a hump shape over time. This finding coincides with the disappearance and reemergence of the Asia-Oceania community observed in Figure \ref{fig1}. In 1995, when the Asia-Oceania community was present, the inter-community trade between Asia-Oceania and America was about 44\% of the intra-community trade within the two communities. In 2002, when the Asia-Oceania community disappeared, the ratio went up to about 51\%. Finally, the ratio went back to about 43\% in 2011, when the Asia-Oceania community was present again.   

\begin{figure}[!t]
\centering
{\includegraphics[width=1\columnwidth]{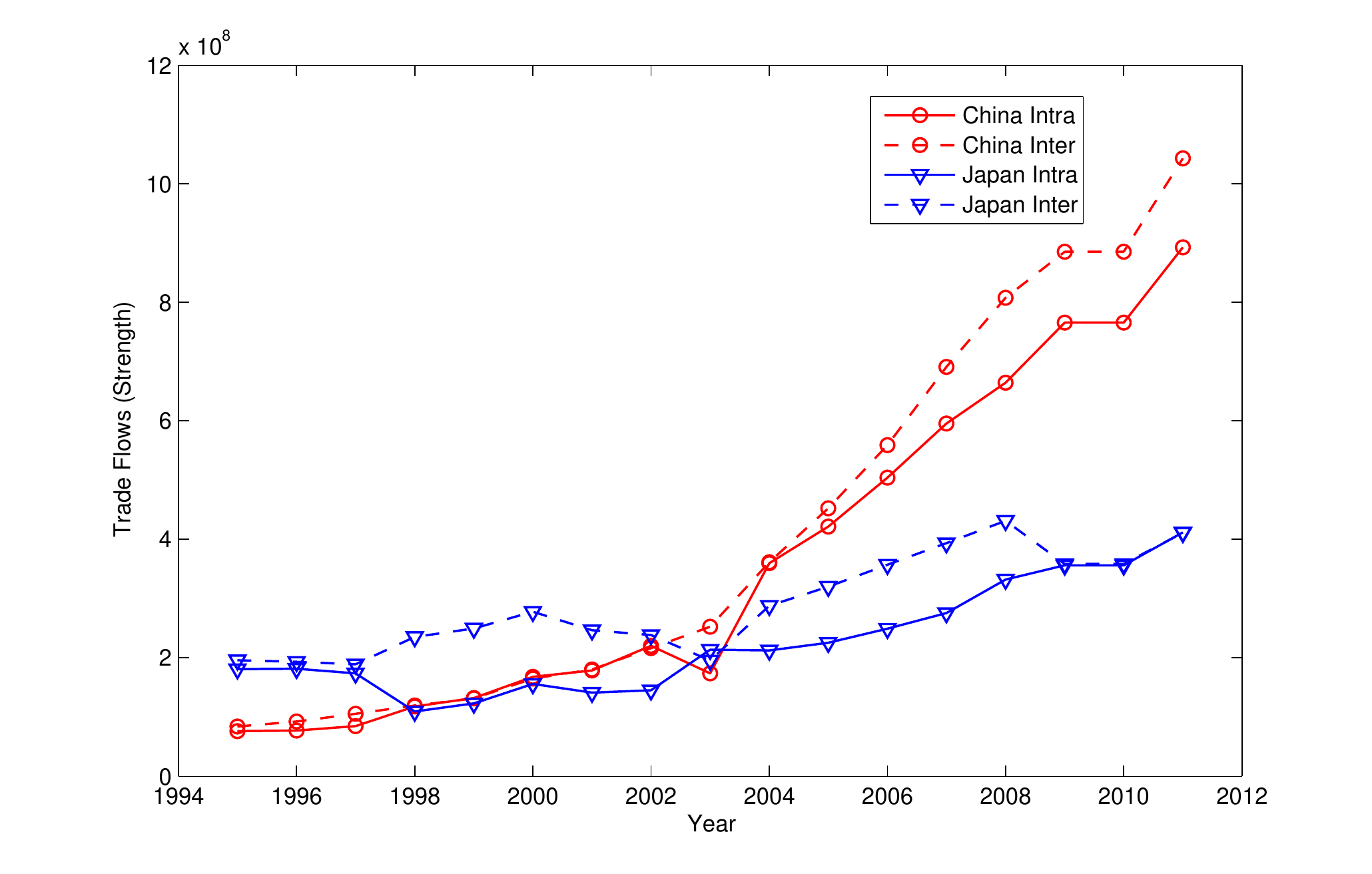}}
\caption{{\bf Intra- and Inter-Community Strength of Japan and China.} We calculate both the inter- and intra-community trade volumes for Japan and China. Japan had more inter-community trade than China before 2003. However, after 2003, China surpassed Japan in terms of both inter- and intra-community trade. This finding coincides with the leadership change from Japan to China observed in Figure \ref{fig1}. Furthermore, for both countries, the intra-community trade follows closely 
to the inter-community trade, which can be viewed as evidence of the intra-community preferential attachment mechanism.} \label{fig4}
\end{figure}

Second, for the intra-community dynamics, we compare the intra-community strength and the inter-community strength between Japan and China. 
As shown in Figure \ref{fig4}, before 2003, Japan always had more inter-community trade than China and had more intra-community trade in the 
beginning and slightly less later. After 2003, China surpassed Japan in terms of both inter- and intra-community trade. This finding coincides with 
the leadership change from Japan to China observed in Figure \ref{fig1}. Also notice that, for both countries, the intra-community trade follows closely 
to the inter-community trade, which can be considered as evidence of the intra-community preferential attachment mechanism.

We also check the regional trade agreements (RTAs) for the intra-community dynamics. Table \ref{table_1} summarizes the effective RTAs signed with 
China during 1995-2011. Only after its accession to WTO in the end of 2001, China started to form RTAs in 2003 and with countries almost exclusively in the Asia-Oceania community.

\begin{table}
\begin{center}
\caption{{\bf China's Effective RTAs.} This table has all the effective RTAs involving China during 1995-2011. (G) stands for Goods and (S) for Services. The data is extracted from the WTO website, \url{http://rtais.wto.org/UI/PublicAllRTAList.aspx}.}
	\resizebox{8.7cm}{!}{
    \begin{tabular}{|l|l|}
    \hline  
     \textbf{RTA Name} & \textbf{Date of Entry into Force} \\ \hline
     China - Hong Kong, China & 29-Jun-2003 \\ \hline
	 China - Macao, China & 17-Oct-2003 \\ \hline
	 ASEAN - China & 01-Jan-2005(G); 01-Jul-2007(S) \\ \hline
	 Chile - China & 01-Oct-2006(G); 01-Aug-2010(S) \\ \hline
	 Pakistan - China & 01-Jul-2007(G); 10-Oct-2009(S) \\ \hline
	 China - New Zealand & 01-Oct-2008 \\ \hline
	 China - Singapore & 01-Jan-2009 \\ \hline
	 Peru - China & 01-Mar-2010 \\ \hline
	 China - Costa Rica & 01-Aug-2011 \\
     \hline
    \end{tabular}
    }
    \label{table_1}
  \end{center}
  \end{table}

Last but not least, it is a well observed fact that the Asia-Oceania community is an active participant of the global production 
chain (or global value chain)~\cite{athukorala2005product,athukorala2006production,baldwin2008spoke}. Therefore, the intra-community 
preference over the nodes with more inter-community strength can be understood as the incentive to have better market access through the regional big player in the global production chain.          

\section{Concluding Remarks} \label{sec:conclusion}

By viewing the international trade system as an interdependent complex network, this paper uses community detection 
and community core detection techniques to examine both the global dynamics, i.e., communities disappear or reemerge, and the 
regional dynamics, i.e., community core changes between community members, in the ITN over the period from 1995 to 2011. We find that the 
Asia-Oceania community has displayed rich dynamics both internally and externally. That is, the Asia-Oceania community was present during 1995-2001 and was 
led by Japan, and then it disappeared and was integrated with the America community during 2002-2004, and finally it reemerged during 2005-2011 and was led by China.   

With a model of weight-driven preferential attachment both inter- and intra-communities, we are able to explain the dynamics observed in the Asia-Oceania community. 
Each period a node will be selected and by chance it may increase its edge weight with an inter-community node (if the edge already exists; otherwise a new edge will be established). 
It will then increase its edge weight with an intra-community node. Outside its own community, the selected node prefers to increase its edge weight with the node with high external strength. 
Inside its own community, it prefers to increase its edge weight with the node with not only high internal strength, but more importantly, high external strength. Our simulation results show that the 
global dynamics, i.e., communities disappear or reemerge can be generated by this model setting. 

In light of the model, the interpretation of the dynamics in the Asia-Oceania community can be that, the community collapsed after China entered the WTO and built strong trade relationships 
with other communities, especially with the external cores, i.e., the United States and Germany, and China became regionally attractive due to the preference of external strength and 
restored the Asia-Oceania community and emerged as the community leader. 

We find some supporting evidence in the trade data. In particular, the behavior of the ratio of the inter-community trade to the intra-community trade between the Asia-Oceania community 
and the America community coincides with the disappearance and reemergence of the Asia-Oceania community. Within the community, China surpassed Japan after 2003 in terms of both inter- 
and intra-community trade. In our simulation, the external strength can only be increased by chance. In reality, however, it can be achieved by a series of strategic moves in trade policy. 
This is evidenced by the surging number of RTAs that China formed since 2003. Moreover, the intra-community preference of the nodes with more inter-community strength can be understood as 
the incentive to have better market access through the regional big player in the global production chain.

\section{Acknowledgments}
Authors acknowledge insightful discussions with Fabio Pammolli and Stefano Schiavo. MR and ZZ acknowledge funding from the MIUR (FIRB project RBFR12BA3Y).
All authors acknowledge support from the FET projects MULTIPLEX 317532 and SIMPOL 610704 and the PNR project CRISIS Lab. 
\bibliographystyle{unsrt}
\bibliography{chinaPaper}

\begin{thebibliography}{10}

\bibitem{ABRMP}
R~Albert and A.-L. Barab\'{a}si.
\newblock {Statistical mechanics of complex networks}.
\newblock {\em Reviews of Modern Physics}, 74(1):47--97, 2002.

\bibitem{GCbook2}
Mark Buchanan, Guido Caldarelli, Paolo {De Los Rios}, and Vendruscolo Michele.
\newblock {\em {Networks in cell biology}}.
\newblock Cambridge University Press, 2010.

\bibitem{SIAM}
M.~E.~J. Newman.
\newblock {The Structure and Function of Complex Networks}.
\newblock {\em SIAM Review}, 45:167--256, 2003.

\bibitem{kitsak2010}
Maksim Kitsak, Massimo Riccaboni, Shlomo Havlin, Fabio Pammolli, and H~Eugene
  Stanley.
\newblock Scale-free models for the structure of business firm networks.
\newblock {\em Physical Review E}, 81(3):036117, 2010.

\bibitem{chessa2013}
Alessandro Chessa, Andrea Morescalchi, Fabio Pammolli, Orion Penner,
  Alexander~M Petersen, and Massimo Riccaboni.
\newblock Is europe evolving toward an integrated research area?
\newblock {\em Science}, 339(6120):650--651, 2013.

\bibitem{nature}
Guido Caldarelli, Alessandro Chessa, Fabio Pammolli, Andrea Gabrielli, and
  Michelangelo Puliga.
\newblock {Reconstructing a credit network}.
\newblock {\em Nature Physics}, 9:125--126, 2013.

\bibitem{hummels2007transportation}
David Hummels.
\newblock Transportation costs and international trade in the second era of
  globalization.
\newblock {\em The Journal of Economic Perspectives}, 21(3):131--154, 2007.

\bibitem{edward1999new}
Mansfield Edward and Helen Milner.
\newblock The new wave of regionalism.
\newblock {\em International organization}, 53(3):589--627, 1999.

\bibitem{chortareas2004trade}
Georgios~E Chortareas and Theodore Pelagidis.
\newblock Trade flows: a facet of regionalism or globalisation?
\newblock {\em Cambridge journal of economics}, 28(2):253--271, 2004.

\bibitem{arribas2009measuring}
Iv{\'a}n Arribas, Francisco P{\'e}rez, and Emili Tortosa-Ausina.
\newblock Measuring globalization of international trade: theory and evidence.
\newblock {\em World Development}, 37(1):127--145, 2009.

\bibitem{serrano2003topology}
Ma~Angeles Serrano and Mari{\'a}n Bogun{\'a}.
\newblock Topology of the world trade web.
\newblock {\em Physical Review E}, 68(1):015101, 2003.

\bibitem{garlaschelli2005structure}
Diego Garlaschelli and Maria~I Loffredo.
\newblock Structure and evolution of the world trade network.
\newblock {\em Physica A: Statistical Mechanics and its Applications},
  355(1):138--144, 2005.

\bibitem{fagiolo2009world}
Giorgio Fagiolo, Javier Reyes, and Stefano Schiavo.
\newblock World-trade web: Topological properties, dynamics, and evolution.
\newblock {\em Physical Review E}, 79(3):036115, 2009.

\bibitem{riccaboni2010}
Massimo Riccaboni and Stefano Schiavo.
\newblock Structure and growth of weighted networks.
\newblock {\em New Journal of Physics}, 12(2):023003, 2010.

\bibitem{de2011world}
Luca De~Benedictis and Lucia Tajoli.
\newblock The world trade network.
\newblock {\em The World Economy}, 34(8):1417--1454, 2011.

\bibitem{riccaboni2013}
Massimo Riccaboni, Alessandro Rossi, and Stefano Schiavo.
\newblock Global networks of trade and bits.
\newblock {\em Journal of Economic Interaction and Coordination}, 8(1):33--56,
  2013.

\bibitem{kim2002longitudinal}
Sangmoon Kim and Eui-Hang Shin.
\newblock A longitudinal analysis of globalization and regionalization in
  international trade: A social network approach.
\newblock {\em Social Forces}, 81(2):445--468, 2002.

\bibitem{tzekina2008evolution}
Irena Tzekina, Karan Danthi, and Daniel~N Rockmore.
\newblock Evolution of community structure in the world trade web.
\newblock {\em The European Physical Journal B}, 63(4):541--545, 2008.

\bibitem{piccardi2012existence}
Carlo Piccardi and Lucia Tajoli.
\newblock Existence and significance of communities in the world trade web.
\newblock {\em Physical Review E}, 85(6):066119, 2012.

\bibitem{reyes2014regional}
Javier Reyes, Rossitza Wooster, and Stuart Shirrell.
\newblock Regional trade agreements and the pattern of trade: A networks
  approach.
\newblock {\em The World Economy}, 2014.

\bibitem{gaulier2010baci}
Guillaume Gaulier and Soledad Zignago.
\newblock Baci: International trade database at the product-level the 1994-2007
  version.
\newblock 2010.

\bibitem{newman2004finding}
Mark~EJ Newman and Michelle Girvan.
\newblock Finding and evaluating community structure in networks.
\newblock {\em Physical review E}, 69(2):026113, 2004.

\bibitem{Note1}
Only with a brief interruption in 1998, when the Asia-Oceania community was
  integrated with the America community. Also, during 1999-2001, while China
  was always a member of the Asia-Oceania community, Japan, Oceania, part of
  the Southeast Asia, and some other Asian economies were integrated with the
  America community.

\bibitem{Note2}
During 1999-2001, when Japan was integrated with America, the Asia-Oceania
  community was led by Hong Kong instead.

\bibitem{fortunato2010}
Santo Fortunato.
\newblock Community detection in graphs.
\newblock {\em Physics Reports}, 486(3–5):75 -- 174, 2010.

\bibitem{fortunato2012}
Santo Fortunato and Marc Barthélemy.
\newblock Resolution limit in community detection.
\newblock {\em Proceedings of the National Academy of Sciences}, 104(1):36--41,
  2007.

\bibitem{deleo2013}
Vincenzo De~Leo, Giovanni Santoboni, Federica Cerina, Mario Mureddu, Luca
  Secchi, and Alessandro Chessa.
\newblock Community core detection in transportation networks.
\newblock {\em Phys. Rev. E}, 88:042810, Oct 2013.

\bibitem{cerina2014}
Federica Cerina, Alessandro Chessa, Fabio Pammolli, and Massimo Riccaboni.
\newblock Network communities within and across borders.
\newblock {\em Scientific Reports}, 4, 2014.

\bibitem{Note3}
See the CEPII website, \protect \url
  {http://www.cepii.fr/CEPII/en/bdd_modele/presentation.asp?id=1}, for further
  information about BACI.

\bibitem{Note4}
See the website of the United Nations Statistics Division, \protect \url
  {https://unstats.un.org/unsd/methods/m49/m49regin.htm}.

\bibitem{Note5}
Countries in Africa and Western Asia don't have consistent community
  memberships over time. Therefore, they are not classified in any of the three
  communities.

\bibitem{Note6}
The results for all years from 1995 to 2011 are in the Appendix.

\bibitem{Note7}
As discussed in Section \ref {sec:intro}, another interesting change in the
  world trade community structure is the emergence of the Arab community after
  2001. This interesting phenomenon deserves further scrutiny in future
  research.

\bibitem{bergstrand1985gravity}
Jeffrey~H Bergstrand.
\newblock The gravity equation in international trade: some microeconomic
  foundations and empirical evidence.
\newblock {\em The review of economics and statistics}, pages 474--481, 1985.

\bibitem{baldwin2006gravity}
Richard Baldwin and Daria Taglioni.
\newblock Gravity for dummies and dummies for gravity equations.
\newblock Technical report, National Bureau of Economic Research, 2006.

\bibitem{carrere2006revisiting}
C{\'e}line Carrere.
\newblock Revisiting the effects of regional trade agreements on trade flows
  with proper specification of the gravity model.
\newblock {\em European Economic Review}, 50(2):223--247, 2006.

\bibitem{Note8}
There exists some related literature to our model. For example, Barrat et
  al.~\cite {barrat2004weighted} and Riccaboni and Schiavo~\cite
  {riccaboni2010} examine the network evolution with dynamic edge weights. Li
  and Maini~\cite {li2005evolving} investigate the network properties with
  preferential attachment both inter- and intra-communities. However, to the
  best of our knowledge, our model is the first attempt to bring the
  preferential attachment mechanism both inter- and intra-communities to the
  context of a weighted network with a fixed number of nodes.

\bibitem{Note9}
In the context of ITN, the communities can be formed, for instance, by
  continents.

\bibitem{Note10}
In the context of the ITN, a high value of $\alpha $ can be interpreted as
  trade barriers such as tariffs, transportation costs, and language
  difference.

\bibitem{athukorala2005product}
Prema-chandra Athukorala.
\newblock Product fragmentation and trade patterns in east asia*.
\newblock {\em Asian Economic Papers}, 4(3):1--27, 2005.

\bibitem{athukorala2006production}
Prema-chandra Athukorala and Nobuaki Yamashita.
\newblock Production fragmentation and trade integration: East asia in a global
  context.
\newblock {\em The North American Journal of Economics and Finance},
  17(3):233--256, 2006.

\bibitem{baldwin2008spoke}
Richard~E Baldwin.
\newblock {\em The Spoke Trap: hub and spoke bilateralism in East Asia}.
\newblock Oxford University Press, 2008.

\bibitem{barrat2004weighted}
Alain Barrat, Marc Barth{\'e}lemy, and Alessandro Vespignani.
\newblock Weighted evolving networks: coupling topology and weight dynamics.
\newblock {\em Physical review letters}, 92(22):228701, 2004.

\bibitem{li2005evolving}
Chunguang Li and Philip~K Maini.
\newblock An evolving network model with community structure.
\newblock {\em Journal of Physics A: Mathematical and General}, 38(45):9741,
  2005.

\end{thebibliography}

\end{document}